\documentclass[aps,pra,preprint]{revtex4}
\usepackage{amsmath}
\usepackage{amsfonts}
\usepackage{amssymb}
\usepackage{graphicx}
\begin{document}

\title{Quantum mechanics without amplitudes}
\author{J. M. A. Figueiredo\\Universidade Federal de Minas Gerais - Dept. de F\'{\i}sica\\Caixa Postal 702 - Belo Horizonte - 30.123-970 - Brazil}
\email[]{josef@fisica.ufmg.br}
\begin{abstract}
Amplitudes are the major logical object in Quantum Theory. Despite this fact
they presents no physical reality and in consequence only observables can be
experimetally checked. We discuss the possibility of a theory of Quantum
Probabilities capable of give full account to quantum phenomena. Advanteges of
this formulation are the evidence of physical processes not described by the
orotodox formulation using amplitudes and the possibility of a full
algoritimization of Quantum Mechanics.
\end{abstract}
\pacs{03.65.Ca, 03.75.-b, 02.50.Ey}
\maketitle
\bigskip\newpage

The difficulties in treating quantum mechanics as a closed theory come from
the fact that its main logical object - the amplitude - has no reality
content, cannot be measured. Because this an interpretation is needed, such as
that of the Copenhagen School, in order to connect properly physics to
formalism. Interpretation introduces additional logical elements, like wave
packet reduction or Schr\"{o}dinger's cat, all also sharing the property of
having no reality content. Recently Di\'{o}si et all \cite{diosi} and Brun
\cite{brun} showed that the measurement process, treated as a quantum open
system, selects paths in the evolution of the amplitudes that are relevant to
another interpretation, that of consistent histories, introduced by Griffiths
in the 80's \cite{griffiths} and by Omnes \cite{omnes} and Gell-Man and Hartle
\cite{gell}. In some sense this interpretation closes the theory because
amplitude histories are selected by chance and this process is
algoritimizable. This means that a physical content may be given to the
process of selecting histories. A code may be writeen, using random numbers,
simulating this process. In this sense amplitude is a non-algoritimizable
object. No physical process can select a path using amplitudes. No dice has a
face selected using amplitudes. In the history interpretation amplitudes are
used so the reality of the evolution of a quantum system still has meaning
only at the measurement time. In this context we can ask about a minimally
open theory, meaning a formulation of quantum theory with a minimum set of
non-algoritimizable objects. It would be desirable that such an approach
eliminate amplitudes from its formalism and treat quantum evolution as a fully
probabilistic process. A difficulty in this case is how to include phase
effects. The possibility of such formulation is the focus of this work. We
will show that quantum mechanics itself provides the elements needed for
establishment of a theory of quantum probabilities and as consequence, new
physics appears, hidden in the amplitude formalism, but fully inserted in its
information content. This possibility is not evident. The vast efforts in
looking for a stochastic justification for Quantum Mechanics have no met with
success. However recently Skorobogatov and Svertilov \cite{slobodov} showed
that quantum probabilities satisfy a non-Markovian extension of the
Chapman-Kolmogorov equation. This put quantum probabilities inside the class
of classical probability theory, a fully algoritimizable object. Although the
noise source is not explicit in this formulation its existence is guaranteed
by the equation itself, as already known in the Fokker-Planck formalism of
Markovian processes.

A theory of quantum probabilities must satisfy two requisites. It must have at
least the same range of amplitude-based calculations and must satisfy an
algoritimizable axiomatic body. The question is to know whether using only
quantum mechanics a complete theory of probabilities can be obtained,
satisfying both requisites. The first one is satisfied by construction since
it is a consequence of calculations using amplitudes and the rules of the
quantum theory. The second one demands that quantum physics be described by
logical objects mappable on processes using classical logic, like tosses of
coins, classical dynamics, stochastic processes or Turing machines. This work
try to shed light on how to address this problem. Since no elements, foreign
to quantum mechanics can be used, it is clear that the second requisite must
be consequence of the first. Thus amplitudes will be treated here as the only
acceptable logical object necessary to drive the formulation of the theory.
For this consider the quantum amplitude written as a time sliced path
integral
\[
\Psi\left(  x,t\right)  =\left(  \frac{m}{2\pi i\hbar\epsilon}\right)
^{\frac{1}{2}}\int\left[  \lim\limits_{N\rightarrow\infty}G_{N}\right]
\Psi_{0}\left(  x_{0}\right)  dx_{0}
\]
where
\[
G_{N}\equiv\left(  \frac{m}{2\pi i\hbar\epsilon}\right)  ^{\frac{N}{2}}%
\int\exp\left(  \frac{i}{\hbar}A_{N}\right)  \prod\limits_{n=1}^{N}dx_{n}
\]
is the Feynman kernel and
\[
A_{N}\equiv\frac{i\epsilon}{\hbar}\sum\limits_{n=1}^{N+1}\left[  \frac
{m}{2\epsilon^{2}}\left(  x_{n}-x_{n-1}\right)  ^{2}-V\left(  x_{n}%
,t_{n}\right)  \right]
\]
the discretized classical action. Here $t_{n}=n\epsilon$ and all paths end at
the space-time point \ $x_{N+1}=x$ and $t=t_{N}$. The probability $\rho\left(
x,t\right)  =\left\vert \Psi\left(  x,t\right)  \right\vert ^{2}$ for the
truncated path integral
\[
\Psi_{N}\left(  x,t\right)  \equiv\left(  \frac{m}{2\pi i\hbar\epsilon
}\right)  ^{\frac{1}{2}}\int G_{N}\Psi_{0}\left(  x_{0}\right)  dx_{0}
\]
is given by
\begin{equation}
\rho_{N}\left(  x,t\right)  =\left\vert \Psi_{N}\left(  x,t\right)
\right\vert ^{2}=\left(  \frac{m}{2\pi\hbar\epsilon}\right)  ^{N+1}\int
G_{N}^{\ast}G_{N}^{\prime}\Psi_{0}^{\ast}\left(  x_{0}\right)  \Psi_{0}\left(
x_{0}^{\prime}\right)  dx_{0}dx_{0}^{\prime} \label{prob}%
\end{equation}
Since only the point $x_{N+1}$ is common to all paths the wave function
evolves freely untill the instant just before we calculate probabilities. At
this moment only those paths that end at $x_{N+1}=x$ are selected and
probabilities depend on the mutual interference of these selected paths. In
terms of non-dimensional variables $x\rightarrow\sqrt{\frac{\hbar\epsilon}{m}%
}x$ and $V\rightarrow\frac{\hbar}{\epsilon}V$ these interfering paths involve
a difference of action given by
\[
\Delta S_{N}=\sum\limits_{n=1}^{N+1}\left[  \frac{1}{2}\left(  x_{n}%
-x_{n-1}\right)  ^{2}-\frac{1}{2}\left(  x_{n}^{\prime}-x_{n-1}^{\prime
}\right)  ^{2}-V\left(  x_{n},t_{n}\right)  +V\left(  x_{n}^{\prime}%
,t_{n}\right)  \right]
\]
In terms of the mean and difference path variables, $\alpha_{n}=\frac{1}%
{2}\left(  x_{n}^{\prime}+x_{n}\right)  $ and $\beta_{n}=x_{n}^{\prime}-x_{n}%
$, this difference presents some novel features. Define the acceleration by
$a_{n}\equiv\alpha_{n+1}+\alpha_{n-1}-2\alpha_{n}$ and the force by
$f_{n}\equiv-\left(  \frac{\partial V}{\partial x}\right)  _{\alpha_{n}}$ in
order to write $S_{N}$ as
\[
S_{N}=\sum\limits_{n=1}^{N}\left[  \left[  a_{n}-f_{n}\right]  \beta_{n}%
+\sum\limits_{k=1}^{\infty}\left[  \left(  \frac{d^{2k+1}V}{dx^{2k+1}}\right)
_{\alpha_{n}}\beta_{n}^{2k+1}\right]  \right]  +\left(  \alpha_{1}-\alpha
_{0}\right)  \beta_{0}
\]
Using these new variables the Markovian character of the Feynman kernel is
broken. This is a consequence of the specific form of the kinetic energy term
that reorders $\Delta S_{N}$ into a non-Markovian chain in the mean path
variable and a local-in-time process in the difference variables. Using
eqn$\left(  \ref{prob}\right)  $ to obtain probabilities we get
\[
\rho_{N}\left(  x,t\right)  =\left(  \frac{1}{2\pi}\right)  ^{N+1}%
\int_{-\infty}^{\infty}\exp\left(  iS_{N}\right)  \Psi_{0}\left(  \alpha
_{0}+\frac{1}{2}\beta_{0}\right)  \Psi_{0}^{\ast}\left(  \alpha_{0}-\frac
{1}{2}\beta_{0}\right)  \prod\limits_{n=0}^{N}d\alpha_{n}d\beta_{n}
\]
For index $n=0$ the above equation can be integrated for $\beta_{0}$ as
\[
\frac{1}{2\pi}\int_{-\infty}^{\infty}\exp\left(  i\left(  \alpha_{1}%
-\alpha_{0}\right)  \beta_{0}\right)  \Psi_{0}^{\ast}\left(  \alpha_{0}%
+\frac{1}{2}\beta_{0}\right)  \Psi_{0}\left(  \alpha_{0}-\frac{1}{2}\beta
_{0}\right)  d\beta_{0}
\]
Writing $\Psi_{0}\left(  x_{0}\right)  =\sqrt{P_{0}\left(  x_{0}\right)  }%
\exp\left(  i\varphi\left(  x_{0}\right)  \right)  $ where the phase $\varphi$
is chosen by an appropriate potential at time zero \cite{ballentine} this
expression becomes
\begin{align}
&  \frac{1}{\pi}\int_{0}^{\infty}\cos\left(  \left(  \alpha_{1}-\alpha
_{0}\right)  \beta_{0}+\varphi\left(  \alpha_{0}+\frac{1}{2}\beta_{0}\right)
-\varphi\left(  \alpha_{0}-\frac{1}{2}\beta_{0}\right)  \right)
\cdot\label{initial}\\
&  \cdot\sqrt{P_{0}\left(  \alpha_{0}+\frac{1}{2}\beta_{0}\right)  }%
\sqrt{P_{0}\left(  \alpha_{0}-\frac{1}{2}\beta_{0}\right)  }d\beta
_{0}\nonumber
\end{align}
This integral is the initial condition $P_{i}\left(  \alpha_{0},v_{0}\right)
$ depending not only on the mean variable $\alpha_{0}$ but also on the
difference velocity $v_{0}\equiv\alpha_{1}-\alpha_{0}$. The resulting initial
probability is the quantum analog of initial conditions for particle position
and momentum in classical mechanics. This is consequence of the non-Markovian
character of the chain since a one-time condition is not sufficient to define
initial conditions for this kind of process. This prescription denies the
possibility of continuous measurement as observed in the quantum Zeno effect.
We interpret eqn$\left(  \ref{initial}\right)  $ as the rule of initial
conditions that defines the theory of quantum probabilities. It displays how
information concerning initial velocities are consistently inserted into the
body of the theory. The nonlocal character is explicit and is responsible for
the persistence of entanglement when an initially entangled state evolves in
time. Given this initial condition the probability is calculated as
\begin{equation}
\rho_{N}\left(  x,t\right)  =\int\prod\limits_{n=0}^{N}d\alpha_{n}P_{i}\left(
\alpha_{0},v_{0}\right)  \prod\limits_{n=1}^{N}\Pi\left(  y_{n},\alpha
_{n}\right)  \delta\left(  y_{n}-a_{n}+f_{n}\right)  dy_{n} \label{chain}%
\end{equation}
where the transition matrix $\Pi$ is given by
\begin{equation}
\Pi\left(  y,\alpha\right)  =\frac{1}{\pi}\int_{0}^{\infty}\cos\left(
y\beta+g\left(  \alpha,\beta\right)  \right)  d\beta\label{transit}%
\end{equation}
and $g\left(  \alpha,\beta\right)  =\sum\limits_{k=1}^{\infty}\left[
\frac{d^{2k+1}V}{d\alpha^{2k+1}}\beta^{2k+1}\right]  $. Formally eqn $\left(
\ref{chain}\right)  $ has the appearance of a non-Markovian stochastic process
\cite{kampen} having $y$ as the noise variable and an associated Langevin
equation $a_{n}=f_{n}+y_{n}$ that exactly resembles the classical case. Notice
that the transition matrix depends on the potential so it changes its
functional form at each point in space and may be even non-stationary,
depending on whether the potential is time-dependent or not. Besides the
trivial case of a free particle only harmonic potentials present a
potential-independent transition matrix with a $\delta$-centered distribution.
In this case the dynamics is purely classical (in the sense of a difference
equation for Newtons's second law) with no fluctuations. Indeterminancies come
from the initial probability only. In the general case the particle sees a
noise whose distribution depends on its dynamical state. Notice that a
feed-back effect between noise and dynamics is not new in physics. In the
Unruh-Davis effect \cite{unruh} an accelerated particle sees a vacuum noise of
blackbody type whose temperature depends on its acceleration. Here noise
probability depends on information furnished by high order derivatives of the
potential at the particle's position. Its functional form is local but, for
analytic potentials, it contain non-local information inserted in the function
$g$. Similar effect was observed in the time domain by Mitchell and Chiao
\cite{chiao} where a analytic electric pulse sent thru a low frequency
bandpass filter presents a negative group velocity due to the information
content of the analytic tail in the pulse. The filter outputs a signal with a
maximum in advance that depends of the whole past history of the pulse.
Similarly here noise distribution depends on information about the potential
existing in the whole space. The physics of this distribution apparently lies
outside the range of quantum theory, at least in the non-relativistic case.
Notwithstanding additional information can be obtained if an appropriate
interpretation of the nature of the stochastic-like chain showed in
eqn$\left(  \ref{chain}\right)  $ is provided. In this the case it is clear
from eqn$\left(  \ref{transit}\right)  $ that $\Pi$ is candidate for a truth
transition probability matrix defining a real stochastic process. Before a
closer analysis of this possibility it is worthwhile to see whether an object
like the mean velocity makes sense. Immediately we obtain
\begin{align*}
\frac{\hbar}{im}\left\langle \Psi^{\ast}\frac{\partial\Psi}{\partial
x}\right\rangle  =&\int_{-\infty}^{\infty}dx\int\prod\limits_{n=0}%
^{N}d\alpha_{n}P_{i}\left(  \alpha_{0},v_{0}\right)  \left(  x-\alpha
_{N}\right)  \prod\limits_{n=1}^{N}\Pi\left(  a_{n}-f_{n},\alpha_{n}\right)
+\\
&\frac{1}{\pi}\int_{-\infty}^{\infty}dx\int\prod\limits_{n=0}^{N}%
d\alpha_{n}P_{i}\left(  \alpha_{0},v_{0}\right)  \times\\
& \prod\limits_{n=1}^{N-1}\Pi\left(  a_{n}-f_{n},\alpha_{n}\right)
\int_{0}^{\infty}\beta\sin\left(  \left(  a_{N}-f_{N}\right)  \beta+g\left(
\alpha,\beta\right)  \right)  d\beta.
\end{align*}
This expression contains an classical mean and a vacuum term with no classical
counterpart. Again a rule, here used to calculate non-local-in-time means is
defined and must also be included in the body of the quantum probability
theory. In this way a protocol may be defined, based only on probabilities and
capable of take into account the full range of predictions already possible
using amplitudes.

Now we will focus on the structure of the stochastic-like chain displayed in
eqn$\left(  \ref{chain}\right)  $ in order to understand to what extent it may
admit an interpretation as a true physical process. It follows from
eqn$\left(  \ref{transit}\right)  $ that the transition matrix is normalized
to one. However, for an arbitrary potential, it is not non-negative. Therefore
$\Pi$ cannot be considered a true transition probability as required by for
stochastic processes, even though the resulting probability $\rho_{N}$ be
non-negative. For the common case of potentials that go to zero at infinity
and that have even symmetry it is easy to see that the difference $V\left(
\alpha+\beta\right)  -V\left(  \alpha-\beta\right)  $ has a range $\lambda$
which may by put conveniently in the form $\lambda=2\pi M$, for some integer
$M$. This means the non-local character of the potential ceases after that
range and the particle is effectively free. In this case the transition matrix
may be separated into two contributions
\[
\Pi\left(  y,\alpha\right)  \simeq\frac{1}{\pi}\int_{0}^{\lambda}\cos\left(
y\beta+g\left(  \alpha,\beta\right)  \right)  d\beta+\delta\left(  y+f\right)
\]
and in consequence it may be decomposed as a difference of two non-negative
matrices given by
\begin{align*}
\Pi &  =T_{+}-T_{-}\\
T_{+}\left(  y,\alpha\right)   &  =\delta\left(  y+f\right)  +\frac{1}{\pi
}\int_{0}^{\lambda}\left(  1+\cos\left(  y\beta+g\left(  \alpha,\beta\right)
\right)  \right)  ^{2}d\beta\\
T_{-}\left(  y,\alpha\right)   &  =\frac{1}{\pi}\int_{0}^{\lambda}\left(
1-\cos\left(  y\beta+g\left(  \alpha,\beta\right)  \right)  \right)
^{2}d\beta
\end{align*}
For non-symmetrical potentials the convergence of the integral defining $\Pi$
is subtle due to its Fresnel-like structure and is dictated by the behavior of
the potential at infinity. A more appropriate treatment must be developed in
this case, where the range $\lambda$ separates the integration interval in two
regions that results in matrices with negative or positive character so a
decomposition similar to the above one follows. Since our focus is stochastic
interpretation of eqn$\left(  \ref{chain}\right)  $ we shall assume that the
decomposition $\Pi=T_{+}-T_{-}$ is always valid. The initial condition $P_{i}$
also presents an analogous decomposition $P_{i}=P_{+}-P_{-}$ where
\[
P_{\pm}=\frac{1}{\pi}\int_{0}^{\infty}\left(  1\pm\cos\left(  \left(
\alpha_{1}-\alpha_{0}\right)  \beta_{0}-\varphi\left(  \alpha_{0}+\beta
_{0}\right)  +\varphi\left(  \alpha_{0}-\beta_{0}\right)  \right)  \right)
^{2}\sqrt{P_{0}\left(  x_{0}+\beta_{0}\right)  }\sqrt{P_{0}\left(  x_{0}%
-\beta_{0}\right)  }d\beta_{0}
\]
In any case both plus and minus signed probabilities are non-negative. Each
minus signed transition probability ascribe a minus sign to its respective
transition matrix so a path with a odd number of minus signed terms adds a
negative contribution to the total probability. We interpret this scenario as
follows. Define a non-correlated stochastic variable $w=\pm1$. To the original
stochastic variable $y$ an internal degree of freedom written as $y^{\left(
w\right)  }$ is assigned. Then at slice $n$ the difference
\[
T_{+}\left(  y^{\left(  w_{+}\right)  },\alpha,t_{n}\right)  -T_{-}\left(
y^{\left(  w_{-}\right)  },\alpha,t_{n}\right)  =w_{+}T_{+}+w_{-}%
T_{-}=\left\langle w\right\rangle
\]
is mean value of this internal stochastic variable, for each realization of
its external value $y$. We call $w$ thereality index, corresponding to
processes with \textquotedblright reality\textquotedblright\ content $\left(
w=1\right)  $ or \textquotedblright anti-reality\textquotedblright\ $\left(
w=-1\right)  $. Each path in space-time has a reality number equal to the
product of reality indices at each slice $n$. If that path has a positive
reality number it adds to the total probability at $x$ and if the reality
number is negative that path adds negatively to the probability at $x$. But at
each time slice transition matrices $T_{+}$ or $T_{-}$ act like transition
matrices in an ordinary stochastic process. In classical processes noise
variables presents only reality content whereas quantum noise variable
presents anti-reality content as well. Thus both classical and quantum
probabilities admit unified description as stochastic processes in the form
\[
\rho_{N}\left(  x,t\right)  =\int\prod\limits_{n=0}^{N}d\alpha_{n}P_{i}\left(
\alpha_{0},v_{0}\right)  \prod\limits_{n=1}^{N}\sum_{w}w_{n}T_{w}\left(
y_{n}^{\left(  w_{n}\right)  },\alpha_{n}\right)  \delta\left(  y_{n}^{\left(
w_{n}\right)  }-a_{n}+f_{n}\right)  dy_{n}^{\left(  w_{n}\right)  }
\]
and this may be generalized to more degrees of freedom, representing a new
class of noise-driven dynamics. This equation proves that quantum mechanics is
algoritimizable. A code may be written describing the particle leaving its
source, with probability $P_{+}$ or $P_{-}$, chosen by a coin toss. A new toss
and a reality content is given to a stochastic transition it suffers according
a classical Langevan equation, with probabilities $T_{+}$ or $T_{-}$ just as
for a classical stochastic processes. In this way a complete path is formed
and the reality number for it is calculated. This means that a quantum
stochastic process has $2^{N}$ times more paths than its classical analogue.
The reality number denies any physical reality assignment to a path because
the trajectories the particle follows is not defined. The resulting histogram
is obtained after a sum of all reality numbers at a given point. Zero
probability at some point means that paths with positive and negative reality
number compensate each other so in fact the role set of trajectories leading
to this point cannot be followed by the particle. This way the reality number
defines a topology in the set of trajectories. A close analogy to the
Consistent Histories approach appears to exist here. The difference is that
here a set of histories of probabilities not amplitudes is considered. Once
probabilities have physical reality a Monte Carlo code may be written and the
evolution of the dynamics is followed completely, as we can do for simulations
using classical dynamics. As stated above a free particle and the harmonic
oscillator evolve without fluctuations. But the assumed initial condition,
which describes the first motion, also has reality indices so the final
probability does involve interference of paths even in this case. It is now
clear that quantum qechanics belongs to the class of the simplest
non-Markovian processes with memory restricted to a single time slice. Since
stationary quantum states do exist a generalization of Markovian theorems for
stationary processes should also exist for the quantum probability case.

The reasoning presented here allows us to assert that quantum mechanics is a
stochastic process with a noise variable presenting an internal degree of
freedom having an internal $Z\left(  1\right)  $ symmetry group. This
stochastic structure is not evident in the amplitude formalism which gives the
probability at some place but cannot describe how it is formed from a more
basic processes. Notice the integral in $T_{+}$ admits a Gaussian
approximation near the origin so it describes process with \textquotedblright
classical analogue\textquotedblright\ while $T_{-}$ is concave at the origin
and is compatible with jump-like probabilities. In higher spatial dimensions
the formulation presents no additional difficulties, at least for scalar
potentials and spinless particles. Here an important issue arises. The
possibility of algoritimization of quantum processes allows modeling of
experiments with small number of particles where amplitude calculations can
give only asymptotic results, for a large number of trials. Accordingly the
transient pattern observed in the shot-by-shot electron diffraction experiment
of Tonomura, Endo, Matsuda and Kawasaki \cite{tonomura} cannot be explained by
amplitude calculations and could be conveniently modeled by our Monte Carlo
code. In this case when the particle hits the screen it gains a gray level,
corresponding to the balance of reality number at that point. Only after a
huge number of trials do we get confidence concerning the real existence of
the particle there, say after a given brightness threshold. Therefore Quantum
Monte Carlo presents a ludicrous pattern, changing in time as in the classical
case yelding nothing real with certainty. Even the histogram is uncertain.
Only the converging histogram, consistent with amplitude calculations is
obtained with probability one. However intermediate histograms give additional
information. If a particle bright it is there with some confidence even before
the complete histogram is obtained. This shows that the probability formalism
possesses more information than the amplitude formalism although it may
present a more complicated mathematical structure in many situations already
successfully treated with amplitude calculations.

The theory treated here is not sufficient to construct a rigorous probability
theory but shows that a formal model may be realizable. Its usefulness may be
limited by operational difficulties in treating practical problems. Anyway it
is possible that the probability formalism be helpful in a better
understanding of non-trivial quantum phenomena like EPR-Bell and delayed
choice experiments because it treats quantum phenomena using only classical
(algoritimizable) logical objects. But more important is the real possibility
of the existence of an appropriate stochastic process describing quantum
phenomena although it appears that its full justification cannot be obtained
within the limits of the simple one-dimensional, spinless and non-relativistic
case assumed here.


\begin{thebibliography}{99}                                                                                               %
\bibitem {diosi}L.Di\'{o}si, N. Gisin, J. Halliwell and I. C. Percival, Phys.
Rev. Lett. \textbf{74}, 203 (1995)

\bibitem {brun}T. A. Brun, Phys. Rev. Lett. \textbf{78}, 1833, (1997)

\bibitem {griffiths}R. Griffiths, J. Stat. Phys. \textbf{36}, 219, (1984)

\bibitem {omnes}R. Omn\`{e}s, J. Stat. Phys. \textbf{53}, 893, (1988)

\bibitem {gell}M. Gell-Mann and J. B. Hartle in \textit{Complexity, Entropy
and the Physics of Information}, ed. W. Zurek, Sante Fe Institute Studies in
the Science of Complexity (Addison-Wesley, Reading, MA, 1990), vol VIII.

\bibitem {kampen}N. G. van Kampen, \textit{Stochastic Processes in Physics and
Chemistry} (North-Holland, Amsterdam, 1992)

\bibitem {unruh}W. G. Unruh, Phys. Rev. D \textbf{14}, 870 (1976); P. C. W.
Davies, J. Phys. A \textbf{8}, 609,(1975).

\bibitem {chiao}M. W. Mitchell and Y. Chiao, Phys. Lett. A \textbf{230}, 133 (1997).

\bibitem {slobodov}G. A. Skorobogatov and S. I. Svertilov, Phys. Rev. A,
\textbf{58}, 3426, (1998).

\bibitem {tonomura}A. Tonomura, J. Endo, T. Matsuda and T. Kawasaki, Am. J.
Phys. \textbf{57}, 117, (1989)

\bibitem {ballentine}Leslie E. Ballentine, Quantum Mechanics, (World
Scientific, Singapore, 1998)
\end{thebibliography}
\end{document}